\documentclass[twocolumn,showpacs,amsmath,amssymb,pra,floatfix,superscriptaddress]{revtex4}

\usepackage{graphicx}
\usepackage{dcolumn}
\usepackage{bm}
\usepackage{bbm}
\usepackage{psfrag}
\usepackage{multirow,amssymb,amsbsy,amsmath}
\usepackage{stmaryrd}
\usepackage{color}

\newcommand{\I}{\textup{i}}

\newcommand{\D}{\textup{d}}

\newcommand{\dd}{\text{d}}
\newcommand{\dod}[2]{\frac{\dd #1}{\dd #2}}

\newcommand{\ddim}{\udelta\kern0.1em}

\newcommand{\beikonst}[2]{\left( #1 \right)_{\kern-0.2em #2}}

\newcommand*{\ket}[1]{\mathopen{|}#1\mathclose{\rangle}}

\newcommand{\comutxt}[2]{[#1,#2]}
\newcommand{\anticom}[2]{\left\{#1,#2\right\}}

\newcommand{\ketbra}[1]{\mathopen{|}#1\mathclose{\rangle}\hspace{-0.25em}\mathopen{\langle}#1\mathclose{|}}
\newcommand{\ketbrap}[2]{\mathopen{|}#1\mathclose{\rangle}\hspace{-0.25em}\mathopen{\langle}#2\mathclose{|}}

\begin{document}

\preprint{APS/123-QED}

%
%

\title{Digital Quantum Simulation with Rydberg Atoms}%

\author{H. Weimer}%
\email{hweimer@cfa.harvard.edu}%
\affiliation{Department of Physics, Harvard University, 17 Oxford Street, Cambridge, MA 02138, USA} %
\affiliation{ITAMP, Harvard-Smithsonian Center for Astrophysics, 60 Garden Street, Cambridge, MA 02138, USA}%
\author{M. M\"uller}%
\affiliation{Institut f\"ur Theoretische Physik der Universit\"at Innsbruck, und Institut f\"ur Quantenoptik und Quanteninformation der \"Osterreichischen Akademie der Wissenschaften, 6020 Innsbruck, Austria}%
\affiliation{Departamento de F\'isica Te\'orica I, Universidad Complutense, 28040 Madrid, Spain}
\author{H. P. B\"uchler}
\affiliation{Institute for Theoretical Physics III, University of Stuttgart, Germany}
\author{I. Lesanovsky}%
\email{igor.lesanovsky@nottingham.ac.uk}%
\affiliation{Midlands Ultracold Atom Research Centre (MUARC), School of Physics and Astronomy, The University of Nottingham, Nottingham, NG7 2RD, United Kingdom}

\date{\today}%

\begin{abstract}
We discuss in detail the implementation of an open-system quantum simulator with Rydberg states of neutral atoms held in an optical lattice. Our scheme allows one to realize both coherent as well as dissipative dynamics of complex spin models involving many-body interactions and constraints. The central building block of the simulation scheme is constituted by a mesoscopic Rydberg gate that permits the entanglement of several atoms in an efficient, robust
and quick protocol. In addition, optical pumping on ancillary atoms provides the dissipative ingredient for engineering the coupling between the system and a tailored environment. As an illustration, we discuss how the simulator enables the simulation of coherent evolution of quantum spin models such as the two-dimensional Heisenberg model and Kitaev's toric code, which involves four-body spin interactions. We moreover show that in principle also the simulation of lattice fermions can be achieved. As an example for controlled dissipative dynamics, we discuss ground state cooling of frustration-free spin Hamiltonians.
\end{abstract}


\maketitle

\section{Introduction}
Simulating the evolution of many-body quantum systems on classical
computers is a complex task, which is believed to be intrinsically
beyond the capabilities of classical computation
\cite{Bernstein1997}. While in the classical case the number of
degrees of freedom scales linearly with the particle number the
computational complexity in treating interacting many-particle quantum
systems increases drastically due to an exponential growth of the
Hilbert space dimension. One way to overcome this difficulty is to
mimic the behavior of the quantum system of interest in an analogue
physical system whose degrees of freedom are well accessible and
controllable \cite{Feynman1982,Lloyd1996}. Initializing this system in
a desired quantum state and measuring its properties after a given
time is then effectively equivalent to having performed a simulation
of the quantum evolution. Such quantum simulators are currently
developed for several physical platforms (see \cite{Buluta2009} for a recent overview), ranging from atomic systems
\cite{Greiner2002,Bartenstein2004,Jordens2008,Schneider2008}, trapped ions \cite{Porras2004,Friedenauer2008,Kim2010,Barreiro2011},
implementations based on nuclear magnetic resonance
\cite{Somaroo1999,Brown2006}, to photonic devices
\cite{Lanyon2010,Kaltenbaek2010}.

\begin{figure}[tb]
  \includegraphics[width=0.95\linewidth]{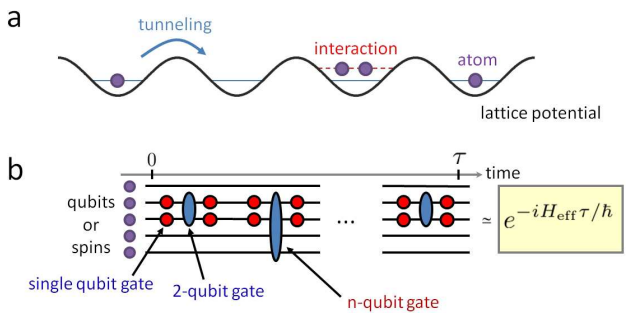}
  \caption{Analogue vs. digital quantum simulator. \textbf{a}: In an analogue simulator the interactions of the simulated model are typically closely related to the physical interactions that underlie the system which implements the  simulator. In the case of cold atoms these are for example optical lattice potentials or short-ranged interatomic interactions, which can be used to simulate Bose and Fermi-Hubbard type Hamiltonians. \textbf{b}:  In the digital case the Hamiltonian evolution is implemented by sequences of quantum gates acting on arrays of qubits, which are e.g. encoded internal states of atoms. The simulated Hamiltonian $H_\mathrm{eff}$ can be vastly different from the interactions governing the underlying physical system. It allows for example the implementation of exotic many-body
  Hamiltonians involving (higher-order) $n$-body interactions, which are very different to those normally encountered in ultracold atomic systems.}
  \label{fig:analogue_vs_digital}
\end{figure}

One paradigmatic system for the simulation of many-body
quantum physics are gases of ultracold alkali atoms trapped in optical
lattices, where the underlying Hamiltonian parameters such as the
hopping rate or onsite interaction strength energies of atoms can be
controlled and tuned externally (see
Fig.~\ref{fig:analogue_vs_digital}) \cite{Jaksch1998}. Since the first observation of a
Mott-insulator to superfluid quantum phase transition
enormous progress has been made in controlling these systems \cite{Bloch2008},
including, very recently, the demonstration of single-site
addressability \cite{Bakr2010,Weitenberg2011}.

Analogue quantum simulators are often purpose built, i.e., they are
particularly well-suited for simulating specific classes of many-body
quantum systems. While the afore-mentioned ultracold atoms in optical
lattices typically realize bosonic or fermionic Hubbard-type
Hamiltonians with short-range interactions, trapped ions for instance
naturally offer the possibility to study interacting spin models. In
general, it is difficult to use analogue quantum simulators for the
study of many-body systems with interactions, that differ considerably
from the natural, physically present (one- and two-body) interactions
underlying the quantum simulator. For the simulation of exotic models
involving certain constraints and higher-order interactions, these
terms are usually created perturbatively. This is typically associated
by fine tuning problems, small effective energy scales and hence slow
dynamics \cite{Albuquerque2008}.

These problems can be overcome by switching to a
circuit based model in a digital, which is sketched in
Fig. \ref{fig:analogue_vs_digital}b. Here, the state of the system is
encoded in qubits. The Hamiltonian time evolution is effectively
created by a sequences of quantum gates which act on these qubits,
i.e. concatenating certain elementary gates will amount to the action
of an effective time-evolution operator whose structure can be
tailored with great flexibility. It has been shown that, provided a
universal set of quantum gates is available, such digital quantum
simulator can be used to simulate the dynamics of any many-body
Hamiltonian with short-range interactions efficiently
\cite{Lloyd1996}. Furthermore it is possible to also efficiently
simulate general (Markovian) dissipative dynamics, by including
dissipative reset operations on ancillary qubits
\cite{Lloyd2001,Bacon2001,Verstraete2009}. Hereby, the engineering of
a controlled coupling of the system to an artificially tailored
environment offers the possibility of realizing dynamics for the
dissipative preparation of entangled states and quantum phases
\cite{Diehl2008}, and closely related, quantum computation based on
dissipation \cite{Verstraete2009}. Recently, we have developed a
physical implementation of an open-system quantum simulator based on
neutral atoms arranged in an optical lattice
\cite{Weimer2010}. Alternative methods for the simulation of spin
systems have recently been discussed: the ground state cooling for the
toric code and a non-abelian topological phase using a single control
atom moving through the lattice and interacting via two-qubit gates
with the system spins has been proposed \cite{Aguado2008}. This method
requires after an initial measurement procedure a second correction
step, which removes the entropy from the system. In addition, the
simulation of the coherent time evolution and the
preparation of thermal states for the toric code using a stroboscopic
method has been discussed by Herdman \textit{et al.}
\cite{Herdman2010}. On the experimental side minimal instances of spin
plaquette models have recently been implemented with trapped ions
\cite{Barreiro2011} and photons \cite{Pachos2009,Ma2011}.

In this work we review and extend such simulation architecture combining the coherent time evolution as well as dissipative terms for interacting spin systems, and in addition provide novel results on the simulation of lattice fermions. The central building block of our simulator is a mesoscopic Rydberg gate which relies on electromagnetically induced transparency (EIT) and the strong and long-ranged interaction of neutral atoms excited to Rydberg states \cite{Muller2009}. We furthermore show that by including optical pumping on ancillary atoms as a dissipative ingredient enables the implementation of open-system many-body dynamics. This dissipative dynamics can be used to perform efficient ground state cooling for a large class of spin models. In general one can expect that the cooling of any frustration-free Hamiltonian can be achieved.

\section{Simulation of coherent dynamics}
\subsection{Setup and general scheme}
In the specific setup we have in mind ultracold atoms are trapped in a deep optical lattice in a Mott-insulator state with a single atom per site. These atoms are used to encode the qubits of our digital simulator in different electronic ground states. The spacing between the lattice sites can be on the order of up to a few micrometers. Experimentally, lattices which grant single
site laser addressability have been demonstrated in Refs. \cite{Nelson2007,Whitlock2009,Bakr2010,Weitenberg2011}.

As sketched in Fig.\ref{fig:analogue_vs_digital}b the temporal evolution in a digital quantum simulator is achieved by concatenating quantum gates. In practice such gates are realized by carefully timed laser pulses and their interplay with state dependent interactions of the trapped atoms.
The general aim is to simulate Hamiltonians of the form
\begin{eqnarray}
  H=\sum_k h_k.
\end{eqnarray}
Here the $h_k$ are quasi-local Hamiltonians that govern the interaction of degrees of freedom located in the vicinity of the $k$-th lattice site. An appropriate sequence of gates will thus implement the time-evolution operator
\begin{eqnarray}
  U(\tau)=\exp\left[-i \tau/\hbar \, \sum_k h_k\right].
\end{eqnarray}
Note that the simulation time $\tau$ is in general different from the real-time $t$. While the idea is simple the practical implementation of such a scheme bears difficulties.

Owed to the finite range of physical interactions, gate operations that can be efficiently implemented are usually quasi-local. The time-evolution operator $U(\tau)$, however, contains highly non-local terms which in case of non-commuting $h_k$ cannot simply be decomposed into products of local ones. In practise one therefore tries to approximate $U(\tau)$ by a sequence of quasi-local gates. Such approximation scheme is provided by the Suzuki-Trotter decomposition \cite{Nielsen2000}. Here one approximately implements the global time-evolution operator over a time step of length $\tau$ by decomposing it into a product of time evolution operators, which correspond to the individual, quasi-local terms $h_k$ of the Hamiltonian:
\begin{eqnarray}
U(\tau)&\approx& \prod_k \exp\left[-i (\tau/\hbar) h_k \right]
\label{eq:trotter}.
\end{eqnarray}
Non-commutativity of terms $h_k$ leads to errors, which are bounded
and can be controlled by the length of the time-step and reduced by
choosing more sophisticated expansion schemes that are a
generalization of eq. (\ref{eq:trotter}) (see, e.g., Ref. \cite{Nielsen2000}).

While in theory any universal set of quantum gates allows the efficient approximation of arbitrary unitaries and therefore the dynamics of arbitrary Hamiltonians with short-range interactions, the experimental realizability strongly demands for an implementation of the time-evolution with few and robust gates. While single-qubit gates are
usually straight-forwardly implemented the challenge lies on creating two-qubit or even many-qubit gates as depicted in Fig. \ref{fig:analogue_vs_digital}b. This key requirement is met by a mesoscopic Rydberg gate which is detailed in the next subsection. It allows the implementation of an entangling multi-qubit gate on a microsecond timescale and with only three laser pulses, independently on the number of atoms involved in the multi-qubit gate. It therefore promises the flexibility, speed and robustness that is necessary for an efficient digital implementation of quantum spin models with exotic interactions.

\subsection{The mesoscopic Rydberg gate}
The purpose of this section is to review the basic properties of the mesoscopic Rydberg gate presented in \cite{Muller2009}, which constitutes the central building block of our envisioned digital quantum simulator. The mechanism underlying the gate operation makes use of a two-photon interference phenomenon known as Electromagnetically Induced Transparency (EIT) \cite{Fleischhauer2005}.
\begin{figure}[tbh]
  \centering
  \includegraphics[width=0.7\linewidth]{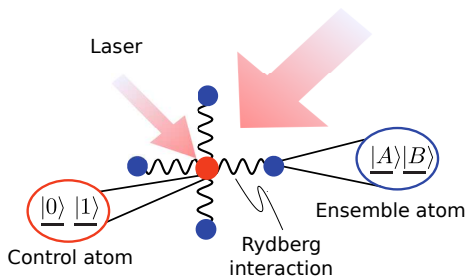}
  \caption{Setup for the mesoscopic quantum gate. A single
    control atom can be addressed independently of $N$ ensemble     atoms. Laser excitations induce a Rydberg interaction between control and ensemble atoms, leading to the realization of a mesoscopic quantum gate.}
  \label{fig:setup}
\end{figure}
The general setup for the implementation of the quantum gate is shown in Fig.~\ref{fig:setup}. We consider a single control atom and $N$ ensemble atoms. For our setup we assume single-site addressability as it has recently been demonstrated by several experimental groups \cite{Nelson2007,Whitlock2009,Bakr2010,Weitenberg2011}. The logical (qubit) states of the control atom are two hyperfine ground states denoted by $\ket{0}$ and $\ket{1}$. The logical states of the ensemble atoms are named $\ket{A}$ and $\ket{B}$. In spite of the different labeling it is in practice not necessary to distinguish between the control and the ensemble atoms - an example for this will be given in Sec. \ref{sec:Kitaev}.

The mesoscopic Rydberg gate uses state-dependent interactions between Rydberg atoms \cite{Jaksch2000,Lukin2001,Brion2007} to realize a Controlled-NOT$^N$ (CNOT$^N$) gate, which is defined by
\begin{equation}
\label{eq:CNOT}
  G=\left|0\right\rangle \!\left\langle 0\right|_{c}\mathop{\otimes}_{i=1}^{N} 1_i+\left|1\right\rangle \!\left\langle 1\right|_{c}\mathop{\otimes}_{i=1}^{N}\sigma^x_i,
\end{equation}
where -- depending on the state of the control qubit -- the state of \textit{all} $N$ target qubits is left unchanged or flipped. Here, $\sigma_i^x \ket{A}_i = \ket{B}_i$ and $\sigma_i^x \ket{B}_i = \ket{A}_i$.

\begin{figure}[tbh]
  \centering
  \includegraphics[width=0.7\linewidth]{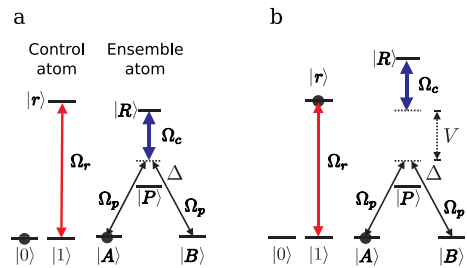}
  \caption{Atomic level structure and external laser couplings for the mesoscopic gate. The states $\ket{1}$ and $\ket{r}$ in the control atom are coupled by a laser with Rabi frequency $\Omega_r$. The weak laser fields $\Omega_p(t)$ drives a Raman transitions from $\ket{A}$ to $\ket{B}$ in the ensemble atoms. (a) For the control atom in $\ket{0}$ the Raman lasers and the strong coupling laser (Rabi frequency $\Omega_c$), coupling $\ket{P}$ to the $\ket{R}$ state, are in two-photon resonance. (b) For the control atom in $\ket{r}$ the Rydberg interaction shifts the $\ket{R}$ level away from the two-photon resonance.}
  \label{fig:pulse}
\end{figure}

To illustrate the underlying mechanism we introduce additional internal levels in both control and ensemble atoms which will be used to physically implement the gate operation (see Fig.~\ref{fig:pulse}): The control atom has an auxiliary Rydberg state $\ket{r}$ that can be coupled to the hyperfine state $\ket{1}$ by a laser.
In the ensemble atoms we employ two additional levels. First, there is a coupling characterized by a time-dependent Rabi frequency $\Omega_p(t)$ between the hyperfine ground states $\ket{A}$ and $\ket{B}$ and an intermediate non-Rydberg $\ket{P}$ level, which still has a low principle quantum number such that interactions with the $\ket{r}$ level of the control atom are negligible. Furthermore, we make use of a Rydberg state $\ket{R}$ in each ensemble atom that is coupled to the intermediate $\ket{P}$ state with a laser of Rabi frequency $\Omega_c$. The external laser fields are chosen such that there is a large detuning $\Delta$ from the $\ket{P}$ level, such that this state is only virtually populated. However, the hyperfine ground states and the Rydberg state are in two-photon resonance.

\begin{figure}[tb]
  \centering
  \includegraphics[width=0.5\linewidth]{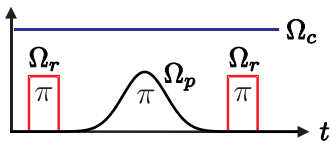}
  \caption{Laser pulse sequence for the mesoscopic gate consisting of an initial $\pi$ pulse on the control atom, an adiabatic Raman transfer in the ensemble atoms, and a second $\pi$ pulse on the control atom.}
  \label{fig:pulse2}
\end{figure}

With this setup the gate operation is performed by a sequence of three laser pulses that is depicted in Fig.~\ref{fig:pulse2}: We start by applying a $\pi$ pulse on the control atom which transforms its qubit state $\alpha\ket{0}+\beta\ket{1}$ to $\alpha\ket{0}+i\beta\ket{r}$.
In the next step we perform a conditional adiabatic Raman transfer in the ensemble atoms from $\ket{A}$ to $\ket{B}$ via the intermediate $\ket{P}$ state. To this end we apply a smoothly varying pulse profile $\Omega_p(t)$, which is chosen such that it realizes an effective $\pi$-pulse between $\ket{A}$ and $\ket{B}$.
Finally, a second $\pi$-pulse is applied to the control atom.

In the following we study the consequences of this pulse sequence on the ensemble atoms in the two cases in which the control atom is in $\ket{0}$ or $\ket{r}$, respectively. The full dynamics then follows by taking the superposition according to the coefficients $\alpha$ and $\beta$. Let us for simplicity first assume that the ensemble atoms do not interact with each other; consequences of non-vanishing interactions will be discussed later. Then, the dynamics of the ensemble atoms reduces to the product of the independent evolution of a single ensemble atom. For a large detuning $\Delta$ we may adiabatically eliminate the $\ket{P}$ level and obtain
the effective Hamiltonian
\begin{align}
H_{\textrm{eff}} = \frac{\hbar\Omega_c^2}{4\Delta}\left[x^2\ketbra{+}
+(1+V)\ketbra{R}+x\left(\ketbrap{+}{R}+\mathrm{h.c.}\right)\right]\label{eq:heffens}\end{align}
Here, $\ket{+} = (\ket{A}+\ket{B})/\sqrt{2}$ is the symmetric
superposition of the two hyperfine ground states and
$x=\sqrt{2}\Omega_p/\Omega_c$ defines the relative strength of the probe laser $\Omega_p$ to the coupling laser $\Omega_c$. Note that during the second laser pulse (Raman transfer) $\Omega_c$ and therefore also $x$ are time-dependent functions. The interaction term $V$ is in fact state-dependent and accounts for the state of the control atom: In an ideal situation we have $V=0$ for the control atom in $\ket{0}$ (see Fig. \ref{fig:pulse}a) while for the control atom in $\ket{r}$ (see Fig. \ref{fig:pulse}b) we have a dominant Rydberg interaction, i.e., $V=\infty$. The antisymmetric state $\ket{-} = (\ket{A}-\ket{B})/\sqrt{2}$ is a zero energy eigenstate of the Hamiltonian. This dark state will thus be unaffected by the dynamics.

Let us now look at the situation in which the control atom is in state $\ket{0}$ (Fig. \ref{fig:pulse}a) and all ensemble atoms are in $\ket{A}$: Here the first and the last $\pi$-pulse shown in Fig. \ref{fig:pulse2} have no effect. We have $V=0$ and we find that Hamiltonian (\ref{eq:heffens}) possesses in addition to $\ket{-}$ a second zero energy dark state,
\begin{equation}
  \left|d\right\rangle =(1+x^{2})^{-1/2}[\left|+\right\rangle-x\left|R\right\rangle ],
\end{equation}
which for $t=0$ corresponds to the $\ket{+}$ state. The only non-zero eigenstate has the energy $E_2 = \frac{\hbar\Omega_c^2}{4\Delta}\left(1+x^2\right)$ and is thus energetically separated from the dark state manifold. During the Raman pulse (see Fig. \ref{fig:pulse2}) the system will adiabatically follow the zero energy dark states for weak coupling lasers with $x\ll 1$ and smooth laser pulse shapes $\Omega_p(t)$. Thus it will follow the dark state $\ket{\bar{d}}=(1/\sqrt{2})\left[\ket{-}+\ket{d}\right]$ which starts and ends in $\ket{A}$. The ensemble is hence effectively transparent for the Raman laser. Imperfections of this adiabatic passage arise from Landau-Zener transitions to the non-zero energy eigenstate \cite{Muller2009}.

In the case of the control atom starting in $\ket{1}$ the first $\pi$-pulse will effectuate a transfer to the Rydberg state $\ket{r}$ and the strong interaction between the Rydberg levels $V$ will change the outcome of the Raman laser sequence. For the sake of simplicity we assume $V=\infty$. Here the Rydberg levels of the ensemble atoms will not take part in the dynamics as they are far off-resonant (see Fig. \ref{fig:pulse}b). Here the time evolution of the ensemble atoms follows the Hamiltonian $H = \hbar\Omega_c^2/(4\Delta)\,x^2\ketbra{+}$. Then, by choosing the pulse shape of $\Omega_p(t)$ ($x\equiv x(t)$) such that $\int x^2(t)\D t = \pi$ the system will undergo the transformation (or Raman transfer)
\begin{eqnarray}
  \ket{-} \to \ket{-}\qquad, \qquad \ket{+} \to -\ket{+}.
\end{eqnarray}
Expressing this transformation in the original states $\ket{A}$ and
$\ket{B}$ results in
\begin{eqnarray}
  \ket{A} \to -\ket{B} \qquad, \qquad \ket{B} \to -\ket{A},
\end{eqnarray}
which is the desired operation up to a trivial phase factor, which can be corrected by choosing suitable phases of the laser fields. Finally, the second $\pi$-pulse de-excites the control atom.

Combining the two scenarios outlined above establishes a way to control a NOT operation on the qubit states of the ensemble atoms conditioned on the state of the control atom, effectively realizing the CNOT$^N$ gate (\ref{eq:CNOT}).
In the next subsections we will make extensive use of this gate when implementing digital quantum simulation schemes for spin models (with many-body interactions).

Before we proceed, however, we want to briefly discuss issues relevant
to the experimental implementation. First of all, we have neglected in
our considerations the interaction among ensemble atoms. This is in
general unjustified since the ensemble atoms are in the course of the gate
sequence excited to Rydberg states which strongly interact. In a
situation in which the control atom is initially in $\ket{1}$ and
hence is excited to a Rydberg state this does not constitute a major
problem since the Rydberg state of the ensemble atoms is shifted far
out of resonance and therefore is not excited. In the opposite
situation (control atom in state $\ket{0}$) the ensemble-ensemble
interaction is however expected to modify the working of the gate
considerably. Indeed, one finds that in this case the state of the
ensemble atoms can no longer be described by a tensor product of dark
states \cite{Muller2009}. Instead one finds that the initial state,
e.g. $\ket{A^N}$ is written as a sum of dark and ''grey'' states which
acquire a dynamical phase shift (with respect to the dark states) that
in the limit of infinite ensemble-ensemble interaction is proportional
to $N \max[x^2(t)]$ \cite{Muller2009}. In order to keep the fidelity
of the gate high this shift has to be kept small. This means the
higher the number of qubits that are to be entangled, the smaller the
ratio $x=\sqrt{2}\Omega_p/\Omega_c$. However, at the same time the Raman transfer has to be carried out at a time much shorter than the lifetime of the atomic Rydberg states, that is typically on the order of $50\,\mu$s. This requires a very strong coupling and also strong Raman lasers. One can show that the gate as presented here can implement an entangling  operation on a timescale of $\sim 1\,\mu$s. In Ref. \cite{Muller2009} we have explicitly shown that fidelities $>99\%$ can be achieved for $N=3$ ensemble atoms located at a distance $\sim 2\mu$m from the control atom. The main limitation to the fidelity is given by the available laser power. In order to achieve high fidelities also for a larger number of ensemble atoms one has to ensure $\sqrt{N} x\ll 1$. This choice suppresses deteriorating effects caused by the interaction among ensemble atoms excited to Rydberg states \cite{Mueller2008,Olmos2011}. Further consequence that arises from gate imperfections on the desired quantum simulation are analyzed towards the end of Sec. \ref{sec:Kitaev}.

\subsection{Digital simulation of spins and fermions}
In the following, we will use the mesoscopic Rydberg gate as building block for the digital quantum simulation \cite{Weimer2010} of spin Hamiltonians. We will at first discuss Kitaev's toric code whose Hamiltonian contains many-body interaction. Despite the seemingly complicated structure this model is rather simple to implement as its Hamiltonian contains no non-commuting parts. The second example concerns the two-dimensional Heisenberg model and as a third example we discuss the digital simulation of lattice fermions. The latter can be mapped onto an effective spin Hamiltonian with six-body interaction terms.

\subsubsection{Kitaev's toric code}\label{sec:Kitaev}
\begin{figure}[tb]
  \includegraphics[width=0.4\linewidth]{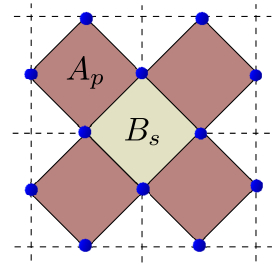}
  \caption{Lattice model for Kitaev's toric code, consisting of two
  sublattices involving plaquette operators $A_p$ and star terms
  $B_s$.}
  \label{fig:storic}
\end{figure}

In Kitaev's toric code model, spins are located on the links of a two-dimensional square lattice and interact via four-body interactions \cite{Kitaev2003}. This model is paradigmatic for a whole class of so-called stabilizer codes \cite{Gottesman1996,Bombin2006}. Its Hamiltonian is given by
\begin{equation}
  H = -E_0 \left(\sum\limits_i A_p^{(i)} + \sum\limits_j B_s^{(j)}\right),
\end{equation}
with the ``plaquette terms'' $A^{(j)}_p =
\sigma_x\sigma_x\sigma_x\sigma_x$ being the product of four Pauli spin
matrices and ``star'' terms $B^{(j)}_s =
\sigma_z\sigma_z\sigma_z\sigma_z$ defined in an analogous manner, see
Fig.~\ref{fig:storic}. In this notation the index $j$ labels the
plaquettes/stars and the four $\sigma$-operators act on the spins
located at the corners of the corresponding square (see
Fig. \ref{fig:storic}). Besides its initially envisioned use as a
quantum memory, this model has recently received considerable
attention in the context of quantum simulation
\cite{Aguado2008,Pachos2009,Nielsen2010}.
The operators $A_p$ and $B_s$ are stabilizer operators with eigenvalues $\pm 1$. The model can be solved exactly, as all plaquette and star terms of the Hamiltonian mutually commute. The global ground state $\ket{\psi}$ is at the same time the ground state of each of the operators $A_p$ and $B_s$:
\begin{eqnarray}
A_p \ket{\psi} &=& \ket{\psi}\nonumber\\
B_s \ket{\psi} &=& \ket{\psi}
\end{eqnarray}
for all plaquettes and stars, respectively. For periodic boundary conditions on a torus the stabilizers satisfy the relations $ \prod_p A_p = 1$ and $\prod_s B_s = 1$. For a system of $N$ atoms there are $N-2$ independent stabilizers. Consequently, the ground state manifold of the system will be four-fold degenerate. For an experimentally more accessible situation of flat two-dimensional lattice structures, the number of holes in the 2D lattice determines the ground state degeneracy.

Excitations of the toric code Hamiltonian can be of two types: violations of the stabilizer constraints of either $A_p$ operators (``magnetic charges") or $B_p$ terms (``electric charges"). They have an energy gap of $4E_0$ as every
violation will affect two stars or plaquettes, respectively. In the following we will illustrate these excitations for the magnetic charges, but due to the
symmetry of the Hamiltonian the situation is identical for the
electric charges.

\begin{figure}[tbh]
\centering
\includegraphics[width=0.95\linewidth]{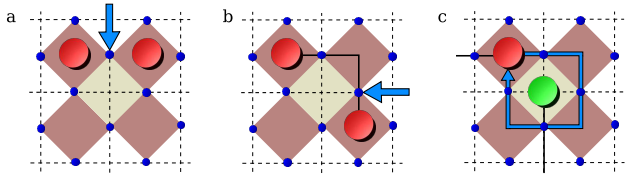}
\caption{Excitations in the toric code. (a) Flipping the spin
  indicated by the arrow will create two magnetic charges on the adjacent plaquettes. (b) Charges can be moved around by flipping further spins on the plaquettes containing the charges. The string operator characterizing the non-local excitation is shown as a solid line. (c) Moving a magnetic charge around an electric charge.}
\label{fig:charges}
\end{figure}

Flipping a single spin will create two magnetic charges located on
adjacent plaquettes, see Fig.~\ref{fig:charges}. By flipping a
different spin on one of the adjacent plaquettes the excitations are
effectively moved. The excitations are no longer quasi-local, but must
be described by a string operator involving the path along which the
charge has been moved. By flipping several spins we also may move a
magnetic charge around an electric charge; due to the
non-commutativity of $\sigma_x$ and $\sigma_z$ the state will
eventually pick up a phase of $\pi$. This behavior shows that the
quasiparticles describing magnetic or electric charges neither have
bosonic nor fermionic character as in both cases one would expect to
recover the identity once the particle had been returned to its
initial position. Hence, one calls such particles with exotic
statistics ``anyons'' related to their potential to pick up ``any''
phase under particle exchange. While the realizations
of systems exhibiting anyonic excitations is interesting in itself,
the anyon dynamics has direct consequences on the ground state cooling
discussed in Sec.~\ref{sec:cool}.

Since the stabilizer operators $A_p$ and $B_s$ mutually commute no Trotter errors occur and the time-evolution operator for a time step $\tau$ is exactly given by
\begin{equation}
  U = \exp(-\I H\tau/\hbar) = \prod_{ps} \exp(\I E_0A_p\tau/\hbar)\exp(\I E_0B_s\tau/\hbar),
\end{equation}
and we need to focus only on the case of a single plaquette or star. The extension to the entire lattice then follows naturally from iterating over all plaquettes and stars. Note that it is possible to parallelize many of these operations by partitioning the system into a few sublattices of piecewise independent atoms.

During each timestep of the digital simulation we want the dynamics of
the plaquette to be governed by the time evolution operator $U_p
=\exp(\I E_0A_pt/\hbar)$. In the regime of
single site addressability, selecting a single plaquette is achieved by focusing the laser pulses which are required for the gate only on the atoms participating in the dynamics of this plaquette. In Ref.~\cite{Weimer2010} a simulation
scheme was suggested where additional ancillary ``control" atoms were
used to effectively mediate four-body interactions between the four
plaquette spins. Here, we propose an alternative approach which works
without ancillary atoms. Instead one of the four plaquette atoms (as shown in Fig. \ref{fig:storic})
will take the role of the control atom $c$. This situation requires a focussing of the
Raman and control lasers exclusively on the remaining three atoms of the plaquette.
This can either be achieved by an appropriate shaping of these laser beams or alternatively by
addressing the three atoms sequentially between the excitation and deexcitation
of the control atom. We can then decompose $U_p$ as
\begin{equation}
  U_p=GU_c^x(\phi)G, \label{eq:kitaev_gate}
\end{equation}
where $G$ describes the mesoscopic Rydberg gate,
\begin{equation}
  G=\left|0\right\rangle \!\left\langle 0\right|_{c}\mathop{\otimes}_{i\ne
  c}1_i+\left|1\right\rangle \!\left\langle 1\right|_{c}\mathop{\otimes}_{i\ne c}\sigma^x_i
\end{equation}
and the single qubit rotation $U_c^x(\phi)=\exp(\I\phi\sigma_c^x)$
imprints a phase of $\phi = E_0t/\hbar$ onto the control atom. The simulation of the star terms $B_s$ follows analogously by performing global $\pi/2$ rotation that interchange $\sigma^x$ and $\sigma^z$. More intuitively, the gate sequence that creates the effective time-evolution under the Hamiltonian $-E\,A_p$ works as follows: The essential step is to realize that the interaction 'sees' only whether a plaquette state has a positive or negative eigenvalue with respect to the operator $A_p$. The actual configuration of the spins is not important. In the first step of the gate sequence (\ref{eq:kitaev_gate}) the information of the state of three plaquette atoms is mapped on the control atom. Now the control atom is manipulated, i.e. a state-dependent rotation is carried out and the initial mapping is reversed. The timescale of this process is essentially set by the speed of the mesoscopic Rydberg gate as
single qubit rotations can be carried out quickly and reliably. Using experimentally realistic parameters, we find an effective interaction strength that can be on the order of several hundred kHz \cite{Weimer2010}.

Let us now briefly investigate the effects of imperfections on the quantum simulator. As mentioned above, a possible error source is the residual Rydberg interaction between the ensemble atoms during the mesoscopic gate, see Ref.~\cite{Muller2009} for a detailed discussion. In the presence of such a coherent error we can write the mesoscopic gate as
\begin{equation}
  G'=|0\rangle\langle0|_{c}\otimes e^{\I\phi
    Q}+|1\rangle\langle1|_{c}\mathop{\otimes}_{i\ne c}\sigma^x_i.
  \label{eq:gateerror}\end{equation}
Here, we have factored out the dependence on the imprinted phase $\phi$ to allow for a consistent expansion. The
perfect gate $G$ then follows in the limit $Q\rightarrow0$, with the Hermitian operator $Q$ acting on the entire plaquette except for the control atom. Expanding $U_p' = G'U_c^x(\phi)G'$ up to first order in
$\phi$ we obtain
\begin{equation}
  U_p' = 1 + 2i\phi Q |0\rangle\langle0|_c + i\phi A_p + O(\phi^2),
\end{equation}
which corresponds to the coherent dynamics under the modified
Hamiltonian $H' = H + (1_c-\sigma^z_c)Q$. Consequently, imperfections can be naturally integrated into the framework of this digital simulator with incoherent errors leading essentially to a finite temperature for the simulated system \cite{Weimer2010b}.

\subsubsection{Heisenberg model}
Let us now discuss a scheme for the simulation of coherent dynamics according to the Heisenberg Hamiltonian
\begin{equation}
\label{eq:Heisenberg_ham}
H = - \frac{1}{2} \sum_{i,j} \left(J_x \sigma_i^x \sigma_j^x +J_y \sigma_i^y \sigma_j^y +J_z \sigma_i^z \sigma_j^z\right) + h \sum_i \sigma_i^z
\end{equation}
Here, $J_x$ ($J_y, J_z$) denotes the coupling strength of $x$-type ($y-, z-$type) spin-spin interactions between neighbouring spins, and $h$ denotes the strength of a single particle term playing the role of an effective magnetic field in $z$-direction acting on all spins.  In contrast to the toric code Hamiltonian discussed above, here not all terms in the Hamiltonian commute. Thus the coherent time evolution has to be realized in a Trotter expansion with small time steps to keep Trotter errors from non-commuting terms small.

Simulation of the dynamics due to the magnetic field term for a small time step $\tau$ is straightforward as $\exp(-i h \tau \sum_i \sigma_i^z)$ can be implemented by a global rotation of all spins, realized e.g. by a corresponding AC Stark shift applied to all atoms. The pairwise interaction terms can be built up from two-atom Rydberg gates $G$ and single-qubit rotations as follows. A small time step $\exp(iJ_x \tau \sigma_i^x \sigma_j^x/2) \equiv \exp(i\theta \sigma_i^x \sigma_j^x/2)$ of pairwise $x$-type interactions between neighbouring spins $i$ and $j$ can be written as
\begin{eqnarray}
&  & \exp(i\theta \sigma_i^x \sigma_j^x/2) \nonumber \\
& = & \exp(-i\pi \sigma_i^y/4) \exp(i\theta\sigma_i^z \sigma_j^x/2) \exp(i\pi \sigma_i^y/2) \nonumber \\
& = & \exp(-i\pi \sigma_i^y/4) \exp(i\theta \sigma_j^x/2)\nonumber \\
&  & \exp( -i \theta \sigma_j^x (1-\sigma_i^z)/2) \exp(i\pi \sigma_i^y/4) \nonumber \\
& = & \exp(-i\pi \sigma_i^y/4) \exp(i\theta \sigma_j^x/2)\nonumber \\
&  & \left[ \ketbra{0}_i \otimes 1_j + \ketbra{1}_i \otimes \exp(-i\theta \sigma_j^x/2)\right]\exp(i\pi \sigma_i^y/4) \nonumber \\ \label{eq:Heisenberg_step}
\end{eqnarray}
Up to single-qubit rotations, this corresponds to an entangling
two-qubit Rydberg gate (term in square brackets), where atom $i$ takes
the role of the control qubit, and atom $j$ undergoes a conditional
spin flip. For $\theta = \pi$ this spin flip takes place with unit
probability and the operation reduces to the gate (\ref{eq:CNOT})
discussed above. To minimize Trotter errors in the simulation, small
values $\theta \ll 1$ are required. This can be readily achieved by
adjusting the pulse length of the second pulse in the pulse sequence
of the gate. For shorter times and / or smaller intensities of this
pulse, the Raman transfer between logical states $\ket{A}_j$ and
$\ket{B}_j$ of the target atom corresponds to only a partial
population transfer, and thereby directly realizes the entangling
operation in the last line of (\ref{eq:Heisenberg_step}). The
implementation of $y$ and $z$ interaction terms in
(\ref{eq:Heisenberg_ham}) can be done by combining the described
procedure with single-qubit rotations on both atoms. We note that for
a fixed set of parameters in the Hamiltonian (\ref{eq:Heisenberg_ham})
it can be beneficial to seek shorter decompositions of the evolution
operator on each pair of sites to reduce the simulation cost in terms
of the required two-qubit entangling gates. Note that in contrast to
previous proposals for the quantum simulation of the Heisenberg model
with cold atoms \cite{Jane2003,Santos2004}, the discussed
implementation with Rydberg atoms can acheive energy scales up to $J_x
\sim 100\,\mathrm{KHz}$, which are limited by the available laser
power and not by the microscopic coupling constants.

\subsubsection{Fermi-Hubbard model in two dimensions}
The Heisenberg model can also be seen as the limiting case of a more
general Hamiltonian that also involves the motion of the particles,
which is the fermionic version of the Hubbard model
\cite{Hubbard1963}. Currently, there is great interest in the quantum
simulation of the Fermi-Hubbard model in two-dimensional systems
because of the expected relations to high-temperature
superconductivity. Most approaches are centered around an analog
simulation using ultracold fermionic atoms in optical lattices
\cite{Jordens2008,Schneider2008}, but the experimental requirements
for reaching the temperature regime of magnetic ordering or even the
regime of $d$-wave superfluidity remain very challenging. Here, we
outline a different approach where lattice fermions are mapped on a
spin Hamiltonian with many-body interactions that can be implemented
using our digital quantum simulator \cite{Abrams1997}.

The Hamiltonian describing the single-band Fermi-Hubbard model for spin $1/2$ particles on a square lattice is given by
\begin{equation}
  H = -t\sum\limits_{<ij>\sigma} c^{\dagger}_{i\sigma}c_{j\sigma} + U\sum\limits_{i}n_{i\uparrow}n_{i\downarrow},
\end{equation}
where $c^\dagger_{i\sigma}$ creates a fermion at site $i$ with spin $\sigma$ and $n_{i\sigma} = c_{i\sigma}^\dagger c_{i\sigma}$ is the corresponding number operator. The parameter $t$ describes a hopping of the fermions to adjacent sites, while $U$ accounts for the interactions of two fermions on the same lattice site. The Heisenberg model discussed above follows in the limit $U\to\infty$ at half filling.

Since we are dealing with fermionic particles, the digital quantum simulator needs to incorporate fermionic statistics. So far, we have discussed how to create a universal quantum simulator for spin interactions. While spin $1/2$ particles have the correct fermionic anticommutator $\anticom{\sigma_i^-}{\sigma_i^+}=1_i$ \emph{on-site},
they have bosonic statistics $\comutxt{\sigma_i^-}{\sigma_j^+}=0$
\emph{off-site}. To overcome this, one has to apply a Jordan-Wigner transformation, which has the form
\begin{eqnarray}
  c_i &=& \mathop{\otimes}_{j=1}^{i-1}\sigma_j^z\sigma_i^-\\
  c_i^\dagger &=& \mathop{\otimes}_{j=1}^{i-1}\sigma_j^z\sigma_i^+\\
  c_i^\dagger c_i &=& \frac{1}{2}(1-\sigma_i^z).
\end{eqnarray}
Here, we have to introduce an enumeration of the sites of the
two-dimensional lattice. This can be done, e.g, by starting in the lower left corner, moving to the lower right corner, go up one site, move to the left, and so on until the entire lattice has been scanned over. Then, the presence of a fermion on site $i$ corresponds to the presence of a down spin. Conversely, an empty site in the fermionic picture translates into a spin up particle. Note that for spin $1/2$ fermions, one has to replace each fermion by two spin $1/2$ particles to encode all four possible combinations on each site.

Let us now look at how the operators in the Fermi-Hubbard Hamiltonian transform. Clearly, the on-site interaction results in interactions of the form $(1-\sigma^z_{i\uparrow})(1-\sigma^z_{i\downarrow})$. On the
other hand, the hopping terms become
\begin{equation}
  c_{i\sigma}^\dagger c_{i+k\sigma} = \sigma^x_{i\sigma}S^\sigma_{i,k}\sigma^x_{i+k\sigma}+ \sigma^y_{i\sigma}S^\sigma_{i,k}\sigma^y_{i+k\sigma}.
\end{equation}
with the string operator $S^\sigma_{i,k} =
\mathop{\otimes}_{j=i+1}^{i+k-1}\sigma^z_{j\sigma}$ acting on
the sites between $i$ and $i+k$. If we consider hopping in the
horizontal direction, we see that within our enumeration scheme the string operators drop out and the resulting spin operators remain local. However, this is not true in the vertical direction. There, we are left with non-vanishing string operators that run across the entire width of the lattice and consequently describe a highly nonlocal interaction.

\begin{figure}[b!]
  \includegraphics[width=0.9\linewidth]{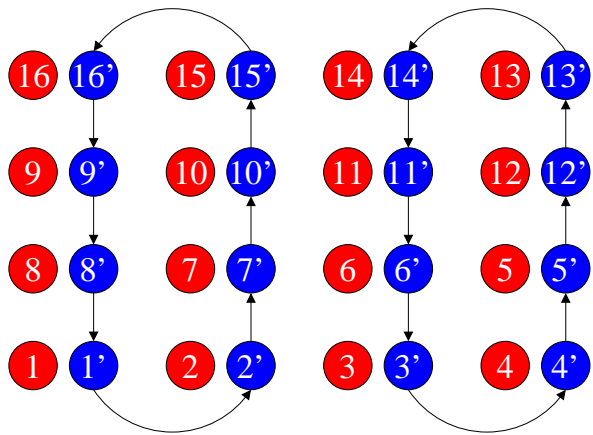}
  \caption{Lattice model for the Fermi-Hubbard model for a single spin degree of freedom. Sites corresponding to the original fermions are colored in red, while the sites corresponding to the auxiliary fermion field are colored in blue. The arrows indicate the graphs along which the projectors $P_{i,j}$ are defined.}
    \label{fig:aux}
\end{figure}

This obstacle can be overcome by introducing auxiliary degrees of freedom as shown in Ref.~\cite{Verstraete2005}. Such degrees of freedom are constituted by a fermion field $d_i$ that is prepared in the ground state of the Hamiltonian
\begin{equation}
  H_{\mathrm{aux}} = -V\sum_{\{i,j\}\sigma}P_{i',j'}P_{j'+1,i'-1},
\end{equation}
with the mutually commuting projectors $P_{i',j'} =
(d_{i'\sigma}+d_{i'\sigma}^\dagger)(d_{j'\sigma}-d_{j'\sigma}^\dagger)$
and $\{i',j'\}$ partitioning the lattice into directed graphs, see Fig.~\ref{fig:aux}. The ground state of $H_{\mathrm{aux}}$ is given by the condition $P_{i',j'}=1$ for all $i',j'$. By replacing the vertical hoppings in the Fermi-Hubbard
Hamiltonian by
\begin{equation}
c_{i\sigma}^\dagger c_{i+k\sigma} \mapsto c_{i\sigma}^\dagger c_{i+k\sigma} P_{i',i'+k}
\end{equation}
it is possible to turn them into local spin operators when the
Jordan-Wigner transformation is applied. The price one has to pay for this decoupling is that the resulting Hamiltonian contains six-body interactions \cite{Verstraete2005}. Including all degrees of freedom, the Hamiltonian is of the form
\begin{widetext}
\begin{align}
H =&  -t\sum_{i,j,\sigma}\left(\sigma^x_{i,j,\sigma}\sigma^x_{i+1,j,\sigma}+\sigma^y_{i,j,\sigma}\sigma^y_{i+1,j,\sigma}\right)\sigma^z_{i',j',\sigma} + t \sum_{i,j,\sigma} \left(\sigma^x_{2i,j,\sigma}\sigma^x_{2i,j+1,\sigma}+\sigma^y_{2i,j,\sigma}\sigma^y_{2i,j+1,\sigma}\right)(-1)^{j+1}\sigma^y_{2i',j',\sigma}\sigma^x_{2i',j'+1,\sigma}\nonumber\\
&+ t \sum_{i,j,\sigma} \left(\sigma^x_{2i+1,j,\sigma}\sigma^x_{2i+1,j+1,\sigma}+\sigma^y_{2i+1,j,\sigma}\sigma^y_{2i+1,j+1,\sigma}\right)(-1)^{j+1}\sigma^x_{2i'+1,j',\sigma}\sigma^y_{2i'+1,j'+1,\sigma}+ \frac{U}{4}\sum\limits_{i,j}(1-\sigma^z_{i,j,\uparrow})(1-\sigma^z_{i,j,\downarrow})\nonumber\\
&+V\sum_{i,j,\sigma}\sigma^z_{2i,2j,\sigma}\sigma^z_{2i+1,2j+1,\sigma}\sigma^x_{2i',2j',\sigma}\sigma^x_{2i'+1,2j',\sigma}\sigma^x_{2i'+1,2j',\sigma}\sigma^x_{2i'+1,2j'+1,\sigma}\nonumber\\
&+V\sum_{i,j,\sigma}\sigma^z_{2i+1,2j+1,\sigma}\sigma^z_{2i,2j+2,\sigma}\sigma^x_{2i',2j'+1,\sigma}\sigma^x_{2i'+1,2j'+1,\sigma}\sigma^x_{2i',2j'+2,\sigma}\sigma^x_{2i'+1,2j'+2,\sigma}\nonumber\\
&+V\sum_{i,j,\sigma}\sigma^z_{2i+1,2j,\sigma}\sigma^z_{2i+2,2j+1,\sigma}\sigma^y_{2i'+1,2j',\sigma}\sigma^y_{2i'+2,2j',\sigma}\sigma^y_{2i'+1,2j'+1,\sigma}\sigma^y_{2i'+2,2j'+1,\sigma}\nonumber\\
&+V\sum_{i,j,\sigma}\sigma^z_{2i+1,2j+2,\sigma}\sigma^z_{2i+2,2j+1,\sigma}\sigma^y_{2i'+1,2j'+1,\sigma}\sigma^y_{2i'+2,2j'+1,\sigma}\sigma^y_{2i'+1,2j'+2,\sigma}\sigma^y_{2i'+2,2j'+2,\sigma},
\end{align}
\end{widetext}
where we have moved to a two-dimensional notation for the lattice.

For an experimental realization one needs to implement four spin $1/2$ particles on every site of the original fermionic model. This could be realized either by stacking up the square lattice in four layers or by being able to address all particles per site individually, e.g., by choosing different hyperfine states. Note that the six-body interactions can again be mapped on the familiar $\prod_i\sigma^x_i$ form by applying local $\pi/2$ rotations on the spins involving $\sigma_i^z$. As an example, let us demonstrate the digital simulation
of the Fermion hopping term
$-t(c_{1,1\uparrow}c_{2,1\uparrow}^\dagger+\mathrm{h.c.})$, which transform into
\begin{eqnarray}
h_{1,2} &=& -t\left(\sigma^x_{1,1,\uparrow}\sigma^x_{2,1,\uparrow}+\sigma^y_{1,1,\sigma}\sigma^y_{2,1,\uparrow}\right)\sigma^z_{1',1',\uparrow}\nonumber\\
&=& -t\sigma^x_{1,1,\uparrow}\sigma^x_{2,1,\uparrow}\sigma^z_{1',1',\uparrow}
-t\sigma^y_{1,1,\uparrow}\sigma^y_{2,1,\uparrow}\sigma^z_{1',1',\uparrow}\nonumber\\
&=& h_1+h_2.
\end{eqnarray}
The two terms $h_1$ and $h_2$ can be implemented sequentially
according to the Suzuki-Trotter formula Eq.~(\ref{eq:trotter}). The
time-evolution according to $h_1$ can be simulated by the gate
sequence $U_1 =
U^H_{1',1',\uparrow}GU_c^x(\phi)GU^H_{1',1',\uparrow}$, where $U^H$ is
the Hadamard gate, which interchanges $\sigma^x$ and $\sigma_z$ and
the phase shift $\phi = t\tau/\hbar$ relates the duration of each
timestep $\tau$ to the coupling constant $t$. The term $h_2$ can be
implemented analogously, by including local gates interchanging
$\sigma_x$ and $\sigma_y$ on the sites $1,1,\uparrow$ and
$2,1,\uparrow$.

While the experimental requirements for the implementations of
non-commuting three-body and six-body interaction terms is certainly challenging, this example illustrates the principle power of a digital quantum simulator. By using specially tailored many-body gates it might be possible to group several terms of the Hamiltonian together and thus reduce the number of required gate operations.

\section{Cooling into many-body ground states}

So far, we have discussed the coherent simulation of spin and fermion lattice models. What we have left aside is the preparation of the initial state for this simulated dynamics. Typically, one is not interested in dynamics of arbitrary initial states, but rather in the behavior of a certain class of states, often those with low energy with respect to the simulated Hamiltonian. One possibility would be to adiabatically follow the ground state from an experimentally
accessible initial state; for example, in analog simulation of lattice
models with cold atoms, the system is first cooled to quantum
degeneracy, with the lattice being ramped up adiabatically
afterwards. Here, we follow a different route: we engineer the
interaction with a dissipative environment in such a way that the resulting dynamics cools the system into the many-body ground state of the Hamiltonian of interest
\cite{Diehl2008,Kraus2008,Verstraete2009,Weimer2010,Aguado2008,Herdman2010}. With such a dissipative element the dynamics is no longer unitary, but can be described by a quantum master equation for the system density operator $\rho$, which is of the form
\begin{equation}
\dod{}{t}{\rho}=-\frac{i}{\hbar}\left[H,\rho\right]+\sum_i\gamma_i \left(c_i\rho c_i^{\dagger}-\frac{1}{2}\anticom{c_i^{\dagger}c_i}{\rho}\right),
\end{equation}
and where the rates $\gamma_i$ control the strength of the
dissipation. The goal is then to engineer the jump operators $c_i$ in such a way that the only stationary states of the master equation correspond to the groun dstates of the Hamiltonian of interest. Note that this Hamiltonian does not necessarily correspond to the Hamiltonian $H$ generating the coherent dynamics in the master equation. In fact, we will study the case of purely dissipative dynamics with $H=0$ in the following.

\subsection{Kitaev's toric code}

\label{sec:cool}

\begin{figure}[tb]
  \includegraphics[width=0.55\linewidth]{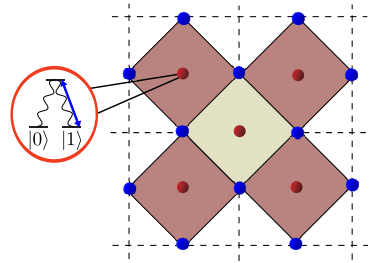}
\caption{Setup for the cooling of the toric
  code. Interstitial control atoms are shown in red, with their   internal level structure allowing optical pumping into the $\ket{0}$ state.}
\label{fig:tcc}
\end{figure}

Let us focus again on the toric code whose coherent simulation was already discussed Sec.~\ref{sec:Kitaev}. In contrast to the implementation of the coherent dynamics we will no longer use one of the atoms of each plaquette or star as the control particle. Instead, we add control atoms to the interstitial spaces of our lattice as shown in Fig.~\ref{fig:tcc}. In our cooling scheme these control atoms will be optically pumped, and therefore provide the coupling to the dissipative environment needed for our quantum state preparation.

As demonstrated in Sec. \ref{sec:Kitaev} the global ground state of the toric
code is at the same time the ground state of each plaquette or
star. Kitaev's toric code falls into the class of frustration-free Hamiltonians, where the energy of each term can be minimized independently. We can therefore perform the cooling to the ground state of each plaquette or star. As in the coherent case we first focus on the case of a single
plaquette $A_p$. Here the local Hamiltonian is given by $h_p =
-E_0\,A_p$ with the spectrum of $A_p$ constituted by two eightfold
degenerate sectors with eigenvalues $+1$ and $-1$. It is therefore
convenient to denote the states by $\ket{\pm 1,\lambda}$, where
$\pm 1$ refers to the eigenvalue of $A_p$ and $\lambda$ labels the different states within the degenerate manifold. The preparation of the ground state sector can be achieved by pumping the plaquette into any superposition or mixture of $+1$ eigenstates. This is realized by choosing a four-body quantum jump operator of the form
\begin{equation}
c_{p}=\frac{1}{2}\sigma_i^z\left(1-A_{p}\right),
\label{eq:jump}
\end{equation}
where $\sigma_i^{z}$ acts on an arbitrary spin $i$ of the four plaquette spins. To understand the action of this jump operator, it is instructive to split it into two parts. First, the ``interrogation'' part $1/2(1-A_p)$ checks whether the system is already in the correct eigenstate. Applied to any $+1$ eigenstate ($A_p = +1$) the jump operator vanishes and the ground state manifold is left unchanged. However, for $A_p = -1$ the second ``pump'' part $\sigma_i^z$ flips the sign of $A_p$ and consequently transforms any state $\ket{-1,\lambda}$ of the excited state manifold directly to the corresponding state $\ket{1,\lambda}$ in the ground state manifold.

The implementation of this jump operator in terms of a gate sequence essentially follows this picture. The auxiliary control atom is initially prepared in the state $\ket{0}_c$. Then, the many-body eigenstate of $A_p = \pm 1$ is mapped onto the control atom by the gate sequence
\begin{equation}
  S = R^y_c(\pi/2)^{-1}GR^y_c(\pi/2),
\end{equation}
where $R_c^y(\pi/2)=\exp(-i\pi\sigma^y_c/4)$ is a local
$\pi/2$ rotation acting on the control atom and $G$ is the mesoscopic Rydberg gate (\ref{eq:CNOT}). The mapping $S$ can be described as
\begin{eqnarray}
  \ket{0}_c\ket{+1,\lambda} &\mapsto& \ket{0}_c\ket{+1,\lambda}\\
  \ket{0}_c\ket{-1,\lambda} &\mapsto& \ket{1}_c\ket{-1,\lambda}
\end{eqnarray}
After this mapping we can therefore conditionally manipulate the many-body states of the plaquette by a conditional operation based on the state of the control atom. We perform a controlled spin flip onto one of the four system spins, given by
\begin{equation}
U^z_i(\theta) = \ketbra{0}_{c}\otimes 1+\ketbra{1}_{c}\otimes \exp(\I \theta \sigma_i^z).
 \end{equation}
Here, the angle $\theta$ controls the probability with which a spin flip from the $-1$ to the $+1$ eigenspace is realized (see below). The two-qubit gate $U^z_i(\theta)$ can be implemented based on the mesocopic Rydberg gate. Finally, we reverse the mapping by applying the inverse gate sequence $S^{-1} (=S)$. Then, we find that the control atom remains in the $\ket{1}$ state every time $U_i^z$ induces a spin flip, i.e.~in general it remains entangled with the four plaquette spins. Consequently, before using the control atom for the next cooling step, optical pumping to the $\ket{0}$ state is required to reinitialize the control atom in $\ket{0}$ such that it factors out from the dynamics of the plaquette spins. It is this dissipative element that provides the necessary ingredient that allows one to remove entropy from the system. For $\theta \ll 1$ we can perform an expansion of the dynamical map describing the evolution of the density operator $\rho$, i.e.,
\begin{equation}
\dod{}{t}\rho=\gamma\left(c_{p}\rho c_{p}^{\dag}-\frac{1}{2}\anticom{c_{p}^{\dag}c_{p}}{\rho}\right)+O(\theta^{3})
\end{equation}
with the jump operators $c_{p}$ given in Eq.~(\ref{eq:jump}) and the cooling rate $\gamma=\theta^{2}/t$.

\begin{figure}[t!b]
  \psfrag{E}[c][c]{$E [E_0]$}
  \psfrag{t1}[r][r]{$\theta = \pi/4$}
  \psfrag{t2}[r][r]{$\theta = \pi/2$}
  \psfrag{t4}[r][r]{$\theta = \pi$}
  \psfrag{t}[c][c]{$t [\tau]$}
  \psfrag{ 0}[c][c]{$0$}
  \psfrag{ 10}[c][c]{$10$}
  \psfrag{ 20}[c][c]{$20$}
  \psfrag{ 30}[c][c]{$30$}
  \psfrag{ 40}[c][c]{$40$}
  \psfrag{-32}[c][c]{$-32$}
  \psfrag{-28}[c][c]{$-28$}
  \psfrag{-24}[c][c]{$-24$}
  \psfrag{-20}[c][c]{$-20$}
  \psfrag{-16}[c][c]{$-16$}
  \includegraphics[width=.95\linewidth]{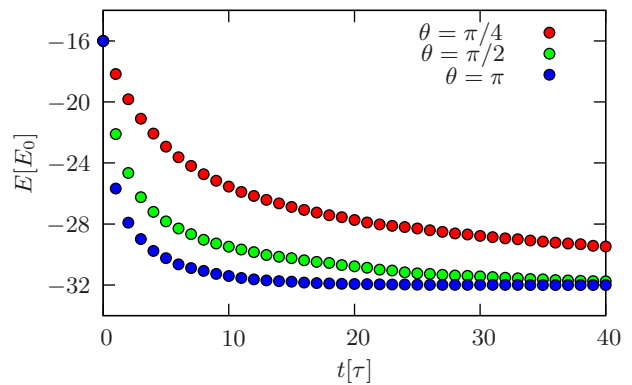}
  \caption{Numerical simulations of the dissipative state preparation
    of the ground state of the toric code for $N=32$ particles. For
    times $t$ large compared to the simulation timestep $\tau$,
    essentially all anyons are removed from the system and the ground
    state energy $E=-NE_0$ is reached asymptotically. By increasing
    the phase shift $\theta$ per timestep, the cooling efficiency can be
    enhanced. Each data point corresponds to an average over 1,000
    realizations.}
  \label{fig:numtc}
\end{figure}
As demonstrated before, this discussion can be generalized to the entire lattice system, with the jump operators for the site terms $B_s$ being given by $c_s = \sigma_i^x(1-B_s)/2$. The cooling process can then be understood in the anyon picture as follows: Each spin flip incoherently moves an anyon to an adjacent plaquette or site, see Fig.~\ref{fig:charges}, and whenever two anyons of the same type meet, they are annihilated and their energy is removed from the system. For
larger values of $\theta$ the dynamics is given by a discrete version of the quantum master equation with the cooling being even more efficient. As can be seen from Fig.~\ref{fig:numtc}, the most efficient preparation of the ground state occurs for $\theta = \pi$. For this value of $\theta$, a dissipative move of an anyon to an adjacent plaquette takes place with unit probability.

\subsection{Frustration-free Hamiltonians}
The above analysis for ground state cooling of the toric code can be extended to a large class of interesting models.
In general, one can design jump operators, where any ground state of a frustration free Hamiltonian is the
unique dark state. With a suitable choice, one would expect the cooling of any initial state into the ground state
of the frustration free Hamiltonian in analogy to the toric code discussed above. An example for a spin liquid phase
at the Roskhar-Kievelson point has been discussed in \cite{Weimer2010}. An important question for the experimental
realization is the efficiency of such ground state cooling: within quantum information theory, the efficient cooling requires that the time evolution of an arbitrary initial state approaches the ground state exponentially with a characteristic cooling rate, which  scales polynomially in the system size.
So far it has been demonstrated that an important subclass of frustration free Hamiltonians, namely stabilizer states, can be cooled efficiently  \cite{Kraus2008, Verstraete2009}. However, it remains an open question whether the ground state of a general frustration free Hamiltonian can be prepared efficiently with dissipative techniques. The toric code discussed above represents an example of a stabilizer state, which exhibits abelian topological order. However, a stabilizer formulation can also be derived for large classes of states exhibiting non-abelian topological order: an important example are the string net condensates \cite{Levin05}. As a consequence, the dissipative ground state cooling illustrated for the toric code above allows one also to efficiently prepare ground states with highly non-abelian topological order. These can serve as the building block for a topological quantum computer.

\section{Conclusion and outlook}

In this work we aimed at discussing the implementation of a digital
quantum simulation architecture using Rydberg atoms in optical
lattices. Our goal was furthermore to outline schemes for the
simulation of coherent and dissipative dynamics corresponding to
(many-body) spin models. Recently, these concepts for
the digital simulation of open-system dynamics have been extended to
systems of trapped ions \cite{Mueller2011}. In a remarkable experiment
\cite{Barreiro2011}, a combination of single- and multi-qubit
(entangling) gates and optical pumping has been used to simulate
coherent four-body spin interactions and dissipative four-qubit
stabilizer pumping, thereby demonstrating in a minimal system of one
plaquette the elementary building blocks required for future
large-scale simulations of Kitaev's toric code and related models.
While the underlying physical interactions of this trapped ions
simulator naturally differ from the van-der-Waals interaction between
Rydberg atoms, the current experiments demonstrate the experimental
feasibility of the digital approach to quantum simulation of
open-system dynamics in interacting many-body systems. Furthermore,
they clearly illustrate the generic effect of how errors in the
employed gates, as discussed in Sec. \ref{sec:Kitaev}, affect the
actual simulated dynamics, and suggest that, ultimately, in future
large-scale fault-tolerant quantum simulation quantum error correction
techniques might have to be incorporated.

For the Rydberg simulator architecture described in the present
work, in principle all individual building blocks have been
experimentally demonstrated individually: The first entangling Rydberg
gates have been recently realized in the laboratory for two atoms held
in optical tweezers \cite{Wilk2010,Isenhower2010}. Important next
steps will be to improve these Rydberg entangling operations and to
incorporate them in larger optical trap arrays and / or optical
lattices, where Mott insulator states involving many lattice sites can
be prepared with high accuracy and single-site addressability has
become available. In combination with the development of efficient
multi-atom entangling Rydberg gates as the one reviewed here
\cite{Muller2009} this appears to be a promising route towards
large-scale quantum simulations of many-body spin models, which lie
beyond the cope of classical simulation capabilities.

\begin{acknowledgments}

H.W. acknowledges support by the National Science Foundation through a grant for the Institute for Theoretical Atomic, Molecular and Optical Physics at Harvard University and Smithsonian Astrophysical Observatory and by a fellowship within the Postdoc Program of the German Academic Exchange Service (DAAD). I.L acknowledges funding by EPSRC. M.M. acknowledges support by the Austrian Science Fund (FOQUS), the European Commission (AQUTE) and the Institute of Quantum Information.

\end{acknowledgments}

\end{document}